\documentclass[final,1p]{elsarticle}

\usepackage{graphicx,url,color,soul}% Include figure files
\usepackage{amssymb}
\biboptions{comma,square}

\journal{Physica A}

\begin{document}
\begin{frontmatter}

\title{Mapping the $q$-voter model: From a single chain to complex networks}

\author[1]{Arkadiusz J\c{e}drzejewski}
\ead{arkadiusz.jedrzejewski@outlook.com}
\author[1]{Katarzyna Sznajd-Weron}
\ead{katarzyna.weron@pwr.edu.pl}
\author[2]{Janusz Szwabi\'nski}
\ead{janusz.szwabinski@ift.uni.wroc.pl}
\address[1]{Department of Theoretical Physics, Wroc{\l}aw University of Technology, Poland}
\address[2]{Institute of Theoretical Physics, University of Wroc{\l}aw, Poland}

\begin{abstract}
We propose and compare six different ways of mapping the modified $q$-voter model to complex networks. Considering square lattices, Barab\'asi-Albert, Watts-Strogatz and real Twitter networks, we ask the question if always a particular choice of the group of influence of a fixed size $q$ leads to different behavior at the macroscopic level. Using Monte Carlo simulations we show that the answer depends on the relative average path length of the network and for real-life topologies the differences between the considered mappings may be negligible.  
\end{abstract}


\end{frontmatter}

\section{Introduction}
Ordering dynamics is not only a classical subject of non-equilibrium statistical physics \cite{Kra:Red:Ben:10,Mor:Liu:Cas:Pas:13}, but also one of the most studied issues in the field of sociophysics \cite{Cas:For:Lor:09,Gal:12,Sen:Cha:13}. It often represents opinion dynamics under the most common type of the social influence, known as conformity. Among many others, models with binary opinions are of particular interest. 

One of the most general models of binary opinion dynamics was introduced by Castellano et al. \cite{Cas:Mun:Pas:09} under the name the $q$-voter model as a simple generalization of the
original voter model~\cite{Cli:Sud:73}. In the proposed model, $q$ randomly picked neighbors (with possible repetitions) influence a voter to change its opinion. If all $q$ neighbors agree, the voter takes their opinion; if they do not have an unanimous opinion, the voter can still flip with probability $\epsilon$. It has been argued that for $q=2$ and $\epsilon=0$ the $q$-voter model coincides with the modified Sznajd model, in which unanimous pair (in one dimension) of the neighboring sites $S_iS_{i+1}$ influences one of two randomly chosen neighbors i.e. $S_{i-1}$ or $S_{i+2}$ \cite{Sla:Szn:Prz:08}.

Following this reasoning in Ref.~\cite{Prz:Szn:Tab:11} we have introduced a modified one-dimensional version of the $q$-voter model, as a natural extension of the original voter and the Sznajd model: a panel of $q$ neighboring spins $S_i,S_{i+1}, . . . ,S_{i+q-1}$ is picked at random. If all $q$ neighbors are in the same state, they influence one of two neighboring spins $S_{i-1}$ or $S_{i+q}$. If not all spins in the $q$-panel are equal then nothing changes in the system. This modification has been later considered in Refs.~\cite{Tim:Pra:14,Tim:Gal:14,Szn:Susz:14}.

For $\epsilon=0$ both formulations of the $q$-voter model seem to be almost identical with the exception of the repetitions possible in the original version \cite{Cas:Mun:Pas:09}. However, there is another difference between formulations \cite{Cas:Mun:Pas:09} and \cite{Sla:Szn:Prz:08}, namely the first belongs to the class of so called inflow and the second to outflow dynamics \cite{Sen:Cha:13,Szn:Kru:06,Roy:Bis:Sen:14}. There was a controversy related to the subject if the inflow and outflow dynamics are equivalent \cite{Sla:Szn:Prz:08,Szn:Kru:06,Beh:Sch:03,Gal:05,Cas:Pas:11}. Recently, it has been shown that they are equivalent for $q=2$ \cite{Cas:Pas:11}, at least in respect to the exit probability. However for larger values of $q$, even in one dimension the situation is not clear \cite{Prz:Szn:Tab:11,Tim:Pra:14,Tim:Gal:14,Roy:Bis:Sen:14,Gal:Mar:11}. Moreover, differences between dynamics in respect to the phase transitions induced by the stochastic noise has not been investigated up till now.

At first glance, it seems to be trivial that choosing a different set of interaction partners will lead to different results on the macroscopic scale. However, as described above, the problem occurred to be not as simple as it seems and gained the attention in the literature. Therefore, one of the aims of this paper is to contribute to the outflow-inflow discussion. 

The second, probably more significant, aim is related to applications of the $q$-voter model in social sciences, because as noted by Macy and Willer  \emph{there was a little effort to provide analysis of how results differ depending on the model designs} \cite{Mac:Wil:02}. Moreover, in respect to social applications one could ask the question -- how to construct the group of influence to create easier an order (consensus) in the system? Another question one could ask -- is the problem relevant for any type of a network or maybe for some networks different types of the group of influence will lead to the same results at the macroscopic scale? Only in the case of a complete graph the definition of the $q$-voter model is straightforward -- since on this topology all spins are neighbors \cite{Nyc:Szn:Cis12,Nyc:Szn:13}, all proposed versions of the $q$-voter model are equivalent.

In the context of opinion dynamics it would be however desirable to consider the models on top of more complex networks, as they are better representations of contact patterns observed in the social systems \cite{alb02,new03}.  There are already several attempts to generalize the $q$-voter model to complex networks \cite{Szn:Susz:14,soo08,mor12,Szn:Szw:Wer:Wer:14}. However, as shown in Ref.~\cite{Sta:Sou:Oli:00}, even in the simple case of transferring the model from 1D chain to a 2D square lattice there is no unique rule of choosing the group of influence. \textbf{Thus, the main goal of this paper is to check, how different ways of picking up the group may impact the macroscopic behavior of the model.
} Specifically, we will focus here on the phase transitions induced by the stochastic noise that represents one type of the social response, known as independence \cite{Nyc:Szn:Cis12,Nyc:Szn:13}.

\section{Model}
Within the modified $q$-voter model we consider a set of $N$ agents called spinsons. This name, being a combination of the words \textit{``spin''} and \textit{``person''}, is used to emphasize that the Ising spins in our model represent persons characterized by only one binary trait (a detailed explanation of this notion  may be found in Ref. \cite{Nyc:Szn:13}).  Each $i$-th spinson has an opinion on some issue that at any given time can take one of two values $S_i=\pm 1, \; i=1,2,\ldots,N$ (``up'' and ``down''). The opinion of a spinson may be changed under the influence of its neighbors according to two different types of the social response \cite{Nai:Mac:Lev:00}: 
\begin{itemize}
\item \textit{Independence} is a particular type of non-conformity. It should be understood as unwillingness to yield to the group pressure. Independence introduces indetermination in the system through an autonomous behavior of the spinsons \cite{Szn:Tab:Tim:11}.
\item \textit{Conformity} is the act of matching spinson's opinion to a group norm. The nature of this interaction is motivated by the psychological observations of the social impact
dating back to Asch~\cite{Asc:55}: if a group of spinson's neighbors unanimously shares an opinion, the spinson will also accept it. 
\end{itemize} 
Other types of the social response are possible as well (see Ref.~\cite{Nyc:Szn:13} for an overview), but the above two are of particular interest for studying opinion dynamics.

We study the model by means of Monte Carlo simulations with a random sequential updating scheme. Each Monte Carlo step consists of $N$ elementary events, each of which may be divided into the following steps: (1) pick a spinson at random, (2) decide with probability $p$, if the spinson will act as independent, (3) if independent, change its opinion with probability $1/2$, (4) if not independent, let the spinson take the opinion of its randomly chosen group of influence, provided the group is unanimous. More details on the dynamic rules of the model may be found in Ref.~\cite{Szn:Szw:Wer:Wer:14}. 
It is worth to stress here the difference between the modified $q$-voter model with independence $p \ge 0$ and the original $q$-voter model with $\epsilon \ge 0$ \cite{Cas:Mun:Pas:09}. In fact one could introduce a generalized model with both parameters $p \ge 0$ and $\epsilon \ge 0$, in which each elementary time step is described by the following algorithm: 
\begin{enumerate}
\item Choose at random one spinson $S_i$ located at site $i$.
\item Decide with probability $p$, if the spinson will act independently.
\item In case of independence, a spinson flips to the opposite state with probability 1/2.
\item In other case (conformity), choose $q$ neighbors of site $i$ (a so called $q$-panel):
\begin{enumerate}
\item If all the $q$ neighbors are in the same state, i.e. $q$-panel is unanimous, the spinson takes the state of the $q$ neighbors.
\item Otherwise, i.e. if $q$-panel is not unanimous, spinson flips with probability $\epsilon$.
\end{enumerate}
\end{enumerate}
Clearly, the original $q$-voter model is a special case of the above algorithm with $p=0$ and the model considered here is a special case with $\epsilon=0$. Note, that in contrast to $p$, parameter $\epsilon$ does not describe the independence. For $p=0$, if only the unanimous $q$-panel exists, the spinson will take its state, which means that it never acts independently.
In consequence, the state with all spinsons in the same state is the absorbing steady state for the original $q$-voter model, whereas it is not for the model considered here unless $p=0$. Therefore, the original $q$-voter model with $p=0$ is not suitable to model e.g. diffusion of innovation, for which the initial state with all spinsons down (unadopted) is  a  typical  one \cite{Szn:Szw:Wer:Wer:14}.

Introduction of a group pressure as one of the rules governing the dynamics assumes some form of interactions between the spinsons. Those interactions are best illustrated as connections between nodes of a graph the spinsons are living on. In its original formulation, the $q$-voter model can be easily investigated on an arbitrary lattice and the definition is clear, since $q$ individuals influencing the voter are chosen with repetitions from the nearest neighborhood of the voter. Therefore, even in one-dimension the parameter $q$ can have an arbitrary value \cite{Cas:Mun:Pas:09}. Although such a definition of the model leads to interesting results from the physical point of view, it seems to be sociologically unreliable. In the modified version, repetitions are forbidden, which is probably more sociologically justified. Moreover, $q$ influencing agents may form panels of different kinds, including structures proposed in \cite{Szn:Susz:14,Szn:Szw:Wer:Wer:14}, which is also different from the original formulation of the model. This in turn allows to investigate the role of the group structure. However, such a modified definition causes ambiguity in mapping the model on an arbitrary graph.

We use here both the Watts-Strogatz \cite{wat98} and the Barab\'asi-Albert networks \cite{bar99} as the underlying topology of spinson-spinson interactions, since they nicely recover the small world property of many real social systems \cite{bar09}. We set $q=4$ for two reasons: (1) to reflect the empirically observed fact that a group of four individuals sharing the same opinion has a high chance to 'convince' the fifth, even if no rational arguments are available \cite{Asc:55,mye13} and (2) to compare our results with those obtained on the square lattice.  

\begin{figure}
\centering
\includegraphics[scale=0.85]{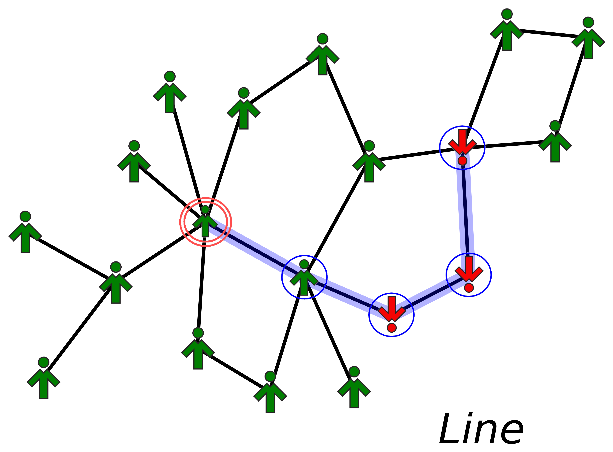} \
\includegraphics[scale=0.85]{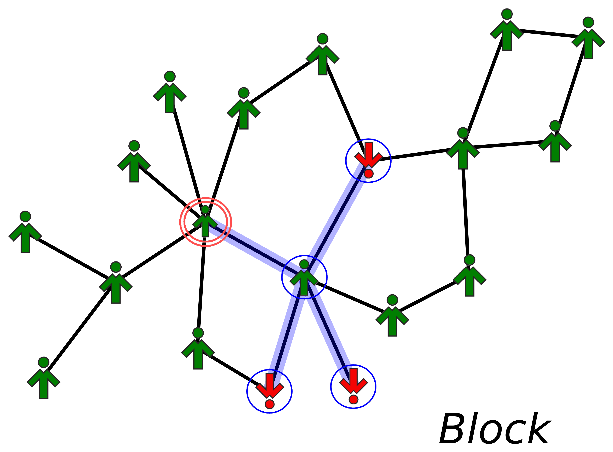} \\
\includegraphics[scale=0.85]{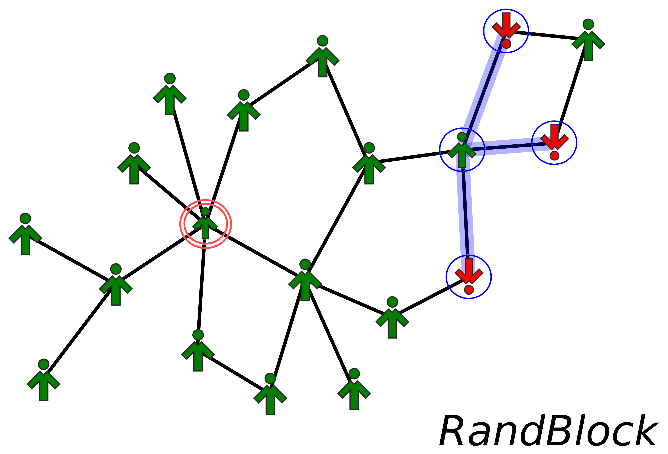} \
\includegraphics[scale=0.85]{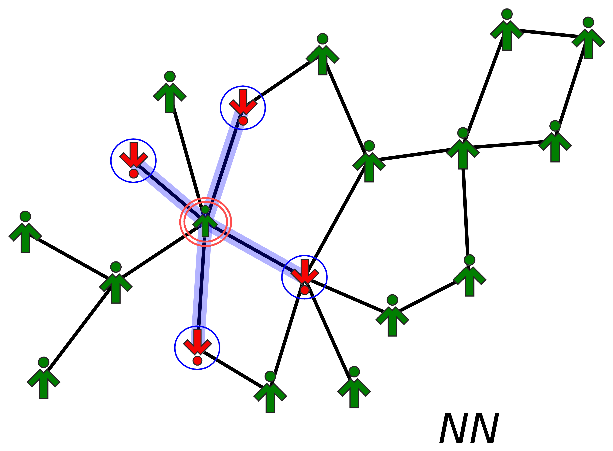} \\
\includegraphics[scale=0.85]{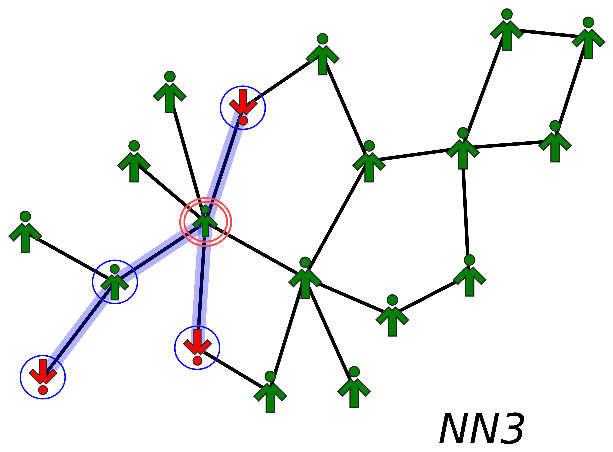} \
\includegraphics[scale=0.85]{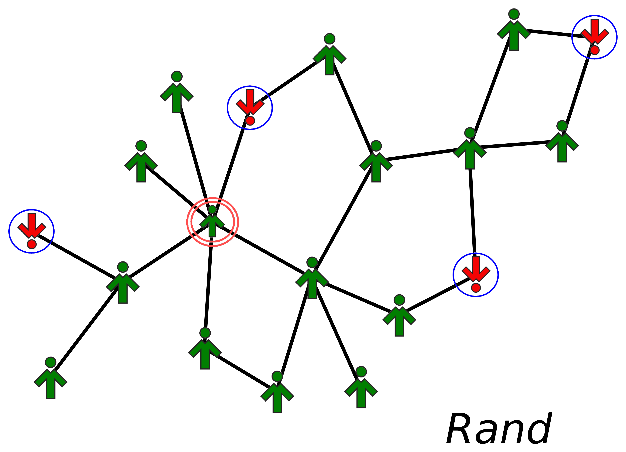} 
\caption{(Color online) Different groups of influence on a complex network. A random spinson is marked with a double red circle, the spinsons in its the group of influence - with a single blue one. Interactions within the group and between the target spinson and the group are represented by thick lines. See text for explanations. \label{influence_group} }
\end{figure}

Chosen from the plethora of possibilities, in Fig.~\ref{influence_group} six different groups of influence on a complex network are schematically shown. We would like to stress here that the choice of precisely such groups is not accidental and is dictated mainly by earlier papers \cite{Cas:Mun:Pas:09,Szn:Susz:14,Szn:Szw:Wer:Wer:14,Nyc:Szn:Cis12}: %They were built in the following way: 

\begin{itemize}
\item $Line$ - after picking up a random target spinson (marked with a double red circle in the figure), we randomly choose one of its neighbors, then one of the neighbors of the neighbor and finally a neighbor of the latter one. All members of the $q$-panel are indicated with a blue circle in the figure. This is the natural generalization of the 1D $q$-voter model and was used e.g. in Ref.~\cite{Szn:Susz:14}.   
\item $Block$ - the group consists of a random neighbor of the target spinson, and three neighbors of the neighbor. This method resembles to some extent the $2\times 2$-block used on square lattices in 2D and was used for instance in  Ref.~\cite{Szn:Szw:Wer:Wer:14}.
\item $NN$ - four randomly chosen nearest neighbors of the target spinson are in the group. This method was used in the original $q$-voter model~\cite{Cas:Mun:Pas:09}.
\item $NN3$ - this is a slight modification of the $NN$ method leading to an extended range of the influence: the group is composed of three randomly chosen nearest neighbors of the target spinson, and a neighbor of one of those nearest neighbors.  We have introduced this method just to investigate the impact of the range of an influence group on system's ability to stay in an ordered state.
\item $RandBlock$ - a spinson and its three neighbors build the the group of influence as in the $Block$ method. However, the block may be located anywhere on the network. This method has been chosen as a reference for the mean-field type approach represented by the next method and similarly as $NN3$ it is aimed to investigate the role of the range of interaction.
\item $Rand$ - the group consists of four randomly chosen spinsons, not necessarily connected with the target spinson. This corresponds to the mean-field approach, for which analytical results on the phase transitions are already known \cite{Nyc:Szn:Cis12}.
\end{itemize}

Note that in case of $RandBlock$ and $Rand$ methods we actually abstract away from the underlying network topology of the model. We expect both methods to be equivalent to the complete graph case if the minimum degree of a node in the network is bigger than or equal to $q=4$ (otherwise the $RandBlock$ may differ slightly from the complete graph, because there will be not always enough spinsons in the neighborhood to build the influence group). Although we use $Rand$ mostly as a benchmark for our simulations, the $RandBlock$ is much more interesting because it corresponds to a situation often encountered in many organizations, in which an informal and unknown network of interactions is over imposed on the given formal communication structure.       

\begin{figure}
\centering
\includegraphics[scale=0.4]{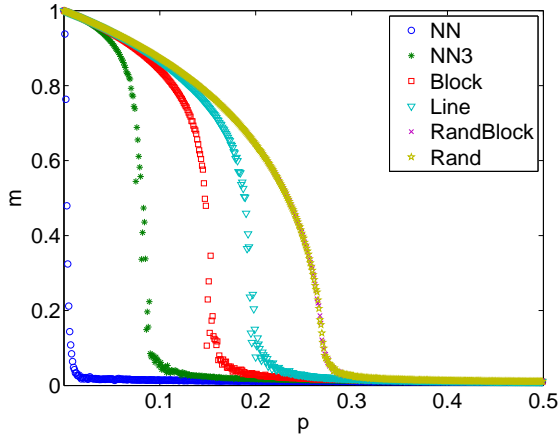} 
\caption{Magnetization $m$ as a function of the independence factor $p$ for different groups of influence on the square lattice $N = 100 \times 100$. As seen, there is no phase transition for NN model on the square lattice and the critical value of $p$ increases with the interaction range, as expected.\label{fig_sq}}
\end{figure}

\begin{figure}
\centering
\includegraphics[scale=0.4]{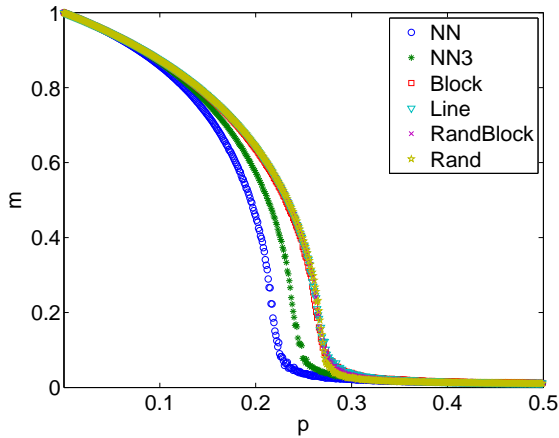} 
\caption{Magnetization $m$ as a function of the independence factor $p$ for six different groups of influence on the Barab\'asi-Albert network of size $N=10^4$ and parameters $M_0=8$ and $M=8$. Four models (Block, Line, RandBlock and Rand) collapse into a single curve and only two (NN and NN3 models) can be distinguished from others.\label{fig_ba}}
\end{figure}

\section{Results}
The main goal of this paper is to answer the questions if and how details at the microscopic level manifest at the macroscopic scale. Among other macroscopic phenomena, phase transitions are certainly the most interesting ones. For the models of opinion dynamics, the most natural order parameter is an average opinion $m$, defined as magnetization i.e. $m=\frac{1}{N} \sum S_i$. It has been shown that in the case of the $q$-voter model the phase transition may be induced by the independence factor $p$ \cite{Szn:Tab:Tim:11,Nyc:Szn:Cis12}. Below the critical value of independence, $p<p_c$, the order parameter $m \ne 0$. For high independence, $p>p_c$, there is a status-quo, i.e. $m=0$. Such results were obtained on the complete graph topology which corresponds to the mean field approach, as well as on the square lattice but only for the one particular choice of the q-panel equivalent to the Sznajd model \cite{Szn:Tab:Tim:11}. In \cite{Szn:Tab:Tim:11} a $2 \times 2$ box of four neighboring spinsons were chosen randomly and influenced one of the 8 neighboring sites of the box. Here we test six different methods described in the previous section (see Fig. \ref{fig_sq}).

It is seen that the critical value of the independence factor strongly depends on procedure of choosing an influence group. The phase transition is observed for all methods except of $NN$, which corresponds to so called inflow dynamics \cite{Roy:Bis:Sen:14}. As expected, $RandBlock$ and $Block$ methods overlap and agree with MFA result found in \cite{Nyc:Szn:Cis12}, i.e. $p_c(q)=(q-1)/(q-1+2^{q-1})$, which for $q=4$ gives $p_c=3/11$. Methods for which the range of interaction is shorter tend to show lower critical value of $p$. This result is very intuitive, since the infinite range of interactions usually corresponds to MFA and gives the largest critical value. 

Results on the Barab\'asi-Albert (BA) network (see Fig. \ref{fig_ba}) are less intuitive. It occurs that for this topology differences between methods are almost negligible. The phase transition is observed for all six models and the critical value of $p$ changes only slightly with method. Four models (Block, Line, RandBlock and Rand) collapse into a single curve and only two (NN and NN3 models) can be distinguished from others. The natural question arises here - why differences between models are clearly visible on the square lattice and are almost negligible in the case of BA? 

It should be recalled that the average path length $l$ for the square lattice increases with the system size $N$ as $l \sim N^{1/2}$, whereas in the case of BA as $l \sim ln N/ln ln N$ \cite{alb02}. It means that for the same system size the average path length is dramatically shorter on BA than on the square lattice. In result the range of interactions on BA is effectively much larger. To check the role of the average path length we have simulated all 6 methods on the Watts-Strogatz network. This topology is particularly convenient because for the fixed network size $N$ it is possible to decrease the average path length $l$ by increasing the rewiring probability $\beta$.

\begin{figure}
\centering
\includegraphics[scale=0.45]{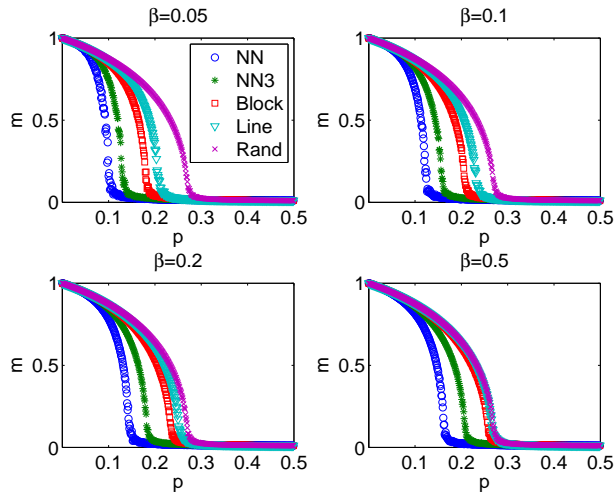} 
\caption{Magnetization $m$ as a function of the independence factor $p$ for different groups of influence on the Watts-Strogatz network of size $N=10^4$ with the average degree $K = 8$ and rewiring parameter $\beta$. The critical value of independence $p_c$ increases with the range of interactions, as expected. However, with increasing $\beta$ differences between models vanish and for large $\beta$ only two models (NN and NN3) can be distinguished from others.\label{fig_ws}}
\end{figure}

\begin{figure}
\centering
\includegraphics[scale=0.45]{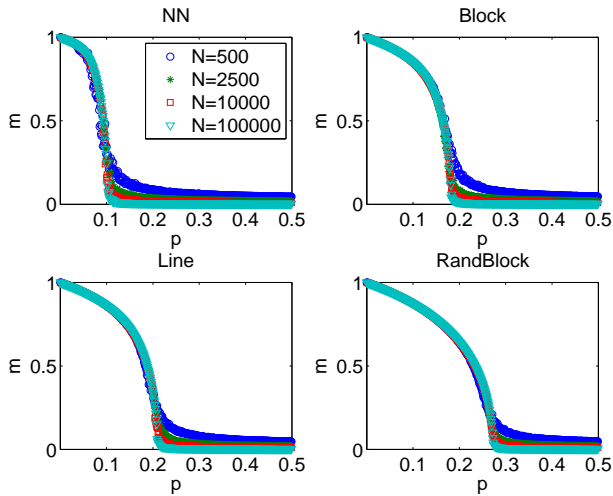} 
\caption{Magnetization $m$ as a function of the independence factor $p$ on the Watts-Strogatz network with the average degree $K = 8$ and rewiring parameter $\beta=0.05$ for several system sizes. Results are not influenced significantly by the system size and the critical value of independence $p_c$ increases with the range of interactions, as expected.\label{fig_wsL}}
\end{figure}

Results for several values of $\beta$ are presented in Fig. \ref{fig_ws}. As $\beta$ increases the critical point $p_c$ shifts towards higher values and simultaneously differences between 4 methods vanish up to a threshold value $\beta=0.5$. Results for larger values of $\beta$ (i.e. $\beta>0.5$; not shown in Fig. \ref{fig_ws}) are identical with those obtained for $\beta=0.5$. To check how results scale with the system size we simulated models on networks of sizes from $N=500$ to $N=10^5$ (see Fig. \ref{fig_wsL}). Surprisingly, results are virtually  independent on the system size. Analogous results has been obtained for the exit probability in one dimensional system with inflow \cite{Roy:Bis:Sen:14} and outflow dynamics \cite{Szn:Susz:14}.

The fact that results do not depend on the system size undermines our predictions that the average path length $l$ itself determines if all mapping methods overlap or not, because the $l$ increases with the system size~\cite{Gos:Bis:Sen:2011}. However, one should probably not look at the path length itself but at the relative path length, which is defined as the average path length of a given network divided by the average path length of a random network of the same size and average degree. Normalizing networks' characteristics by those of the corresponding random graphs is a procedure usually used to compare networks of different sizes~\cite{Fro:Fro:Hol:04,Tel:Joy:Hay:Bur:Lau:2011,Chm:Kli:Thu:2014}. Thus, we will use the relative path length to describe the networks under consideration and to compare them. For example for the Watts-Strogatz network with $k=8$ and $\beta=0.05$ the relative path length is equal $l_{rel}=1.65422863636$ for $N=500$ and $l_{rel}=1.67766421808$ for $N=1000$, i.e. almost size independent. For Barab\'asi-Albert of size $N=500$ it is much shorter i.e. $l_{rel}=0.974$ and almost does not change with the system size. Interestingly, if one consider relative path length $l_{rel}$ for Watts-Strogatz of size $N=500$ and $k=8$ with different $\beta$ it occurs that relative path length approaches $1$ for $\beta=0.5$ and precisely: $l_{rel}(\beta=0.05)=1.65422863636, l_{rel}(\beta=0.1)=1.39668726543, l_{rel}(\beta=0.2)=1.2034321044, l_{rel}(\beta=0.5)=1.02497507194$. This somehow explains why results for BA and WS with $\beta=0.5$ are almost identical.

\begin{figure}
\centering
\includegraphics[scale=0.22]{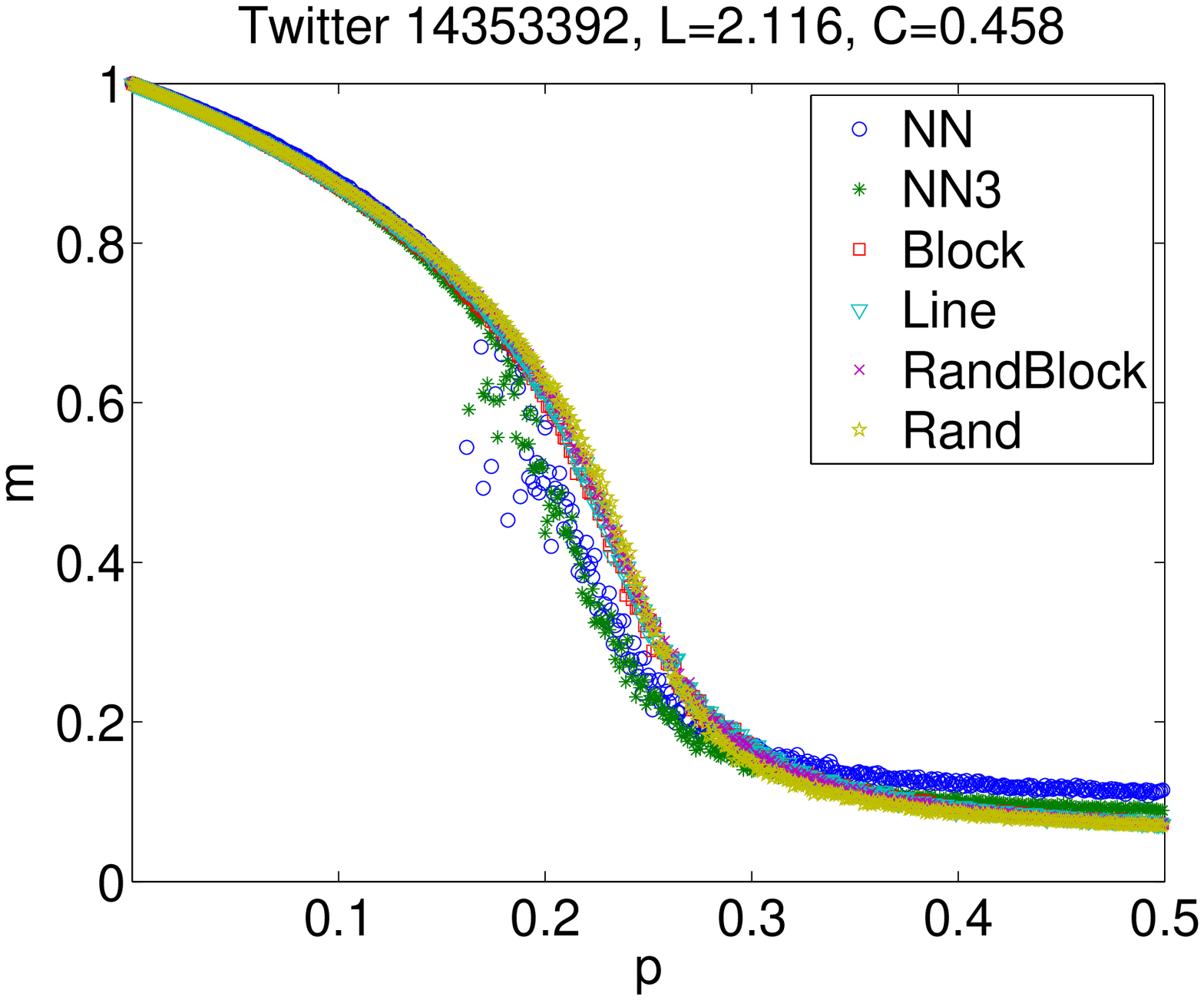} \
\includegraphics[scale=0.22]{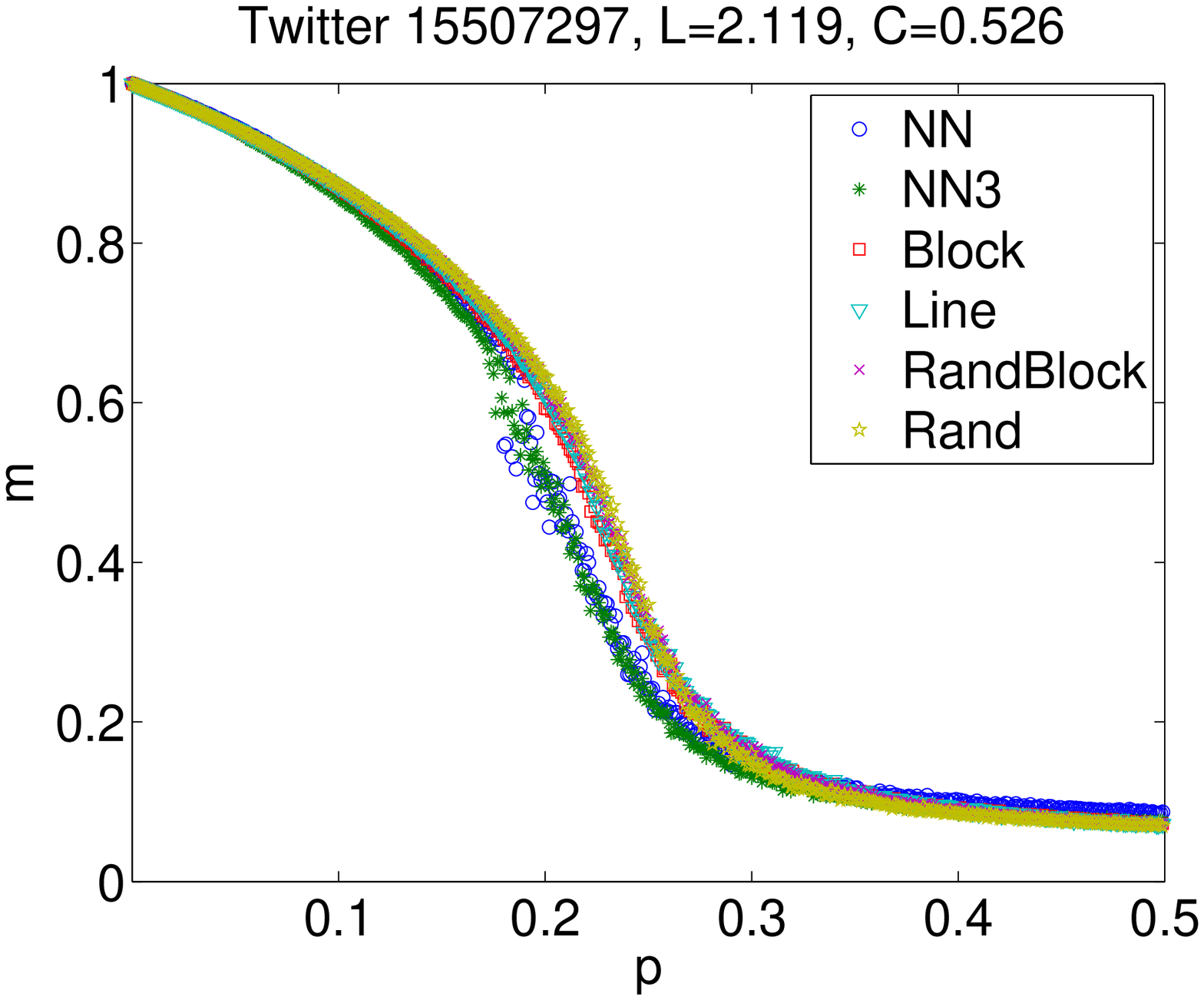} \
\includegraphics[scale=0.22]{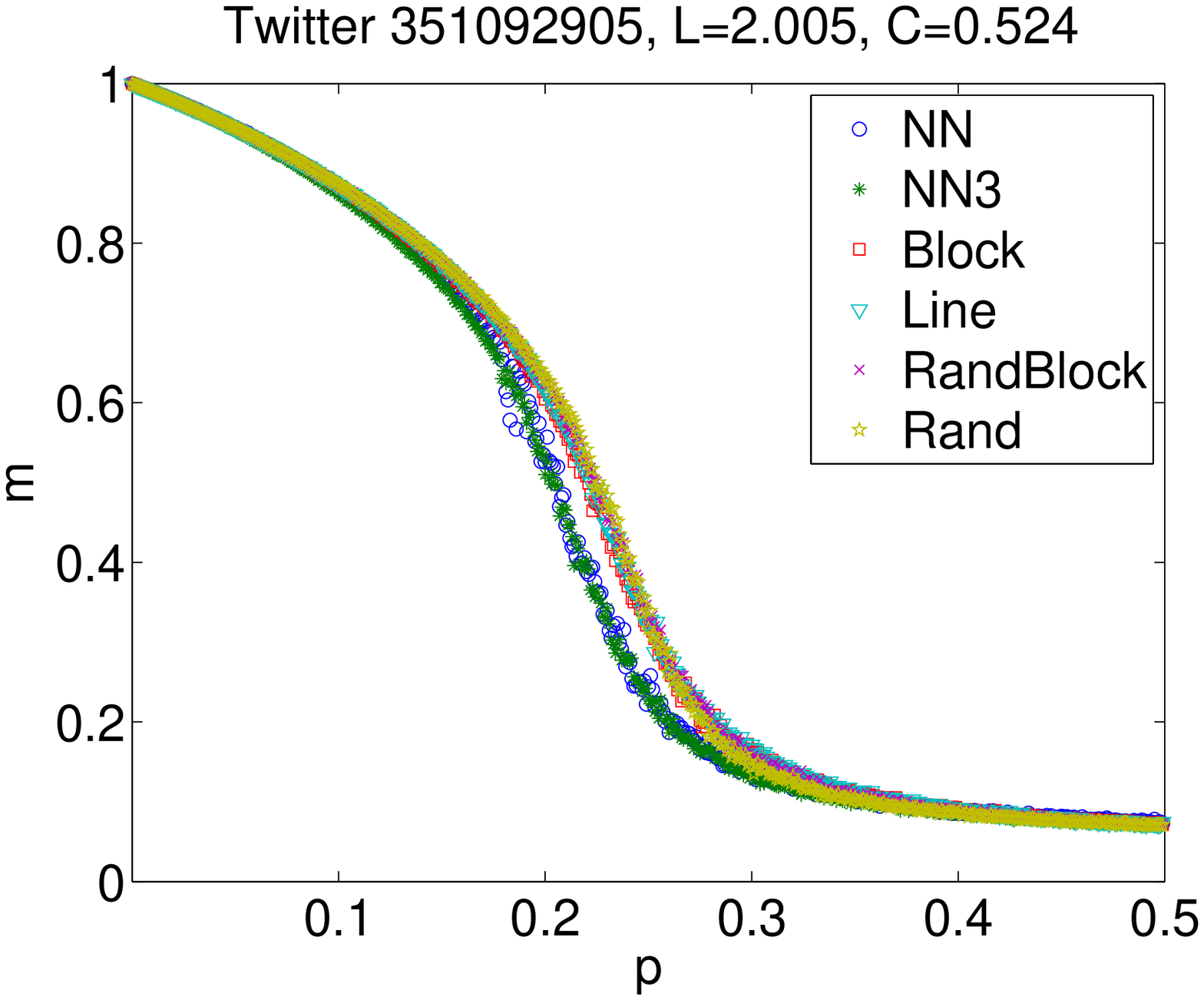} \\
\includegraphics[scale=0.22]{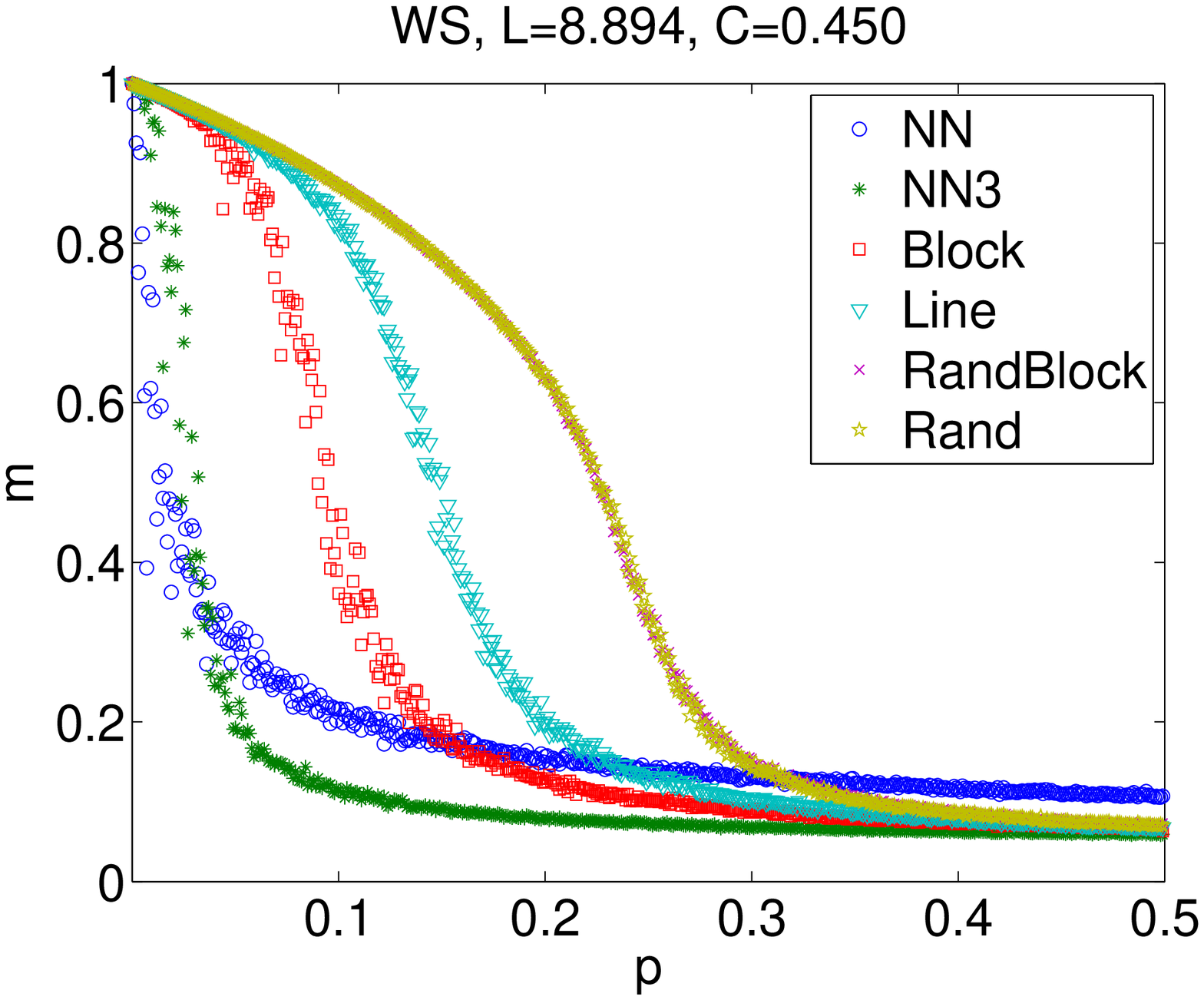} \
\includegraphics[scale=0.22]{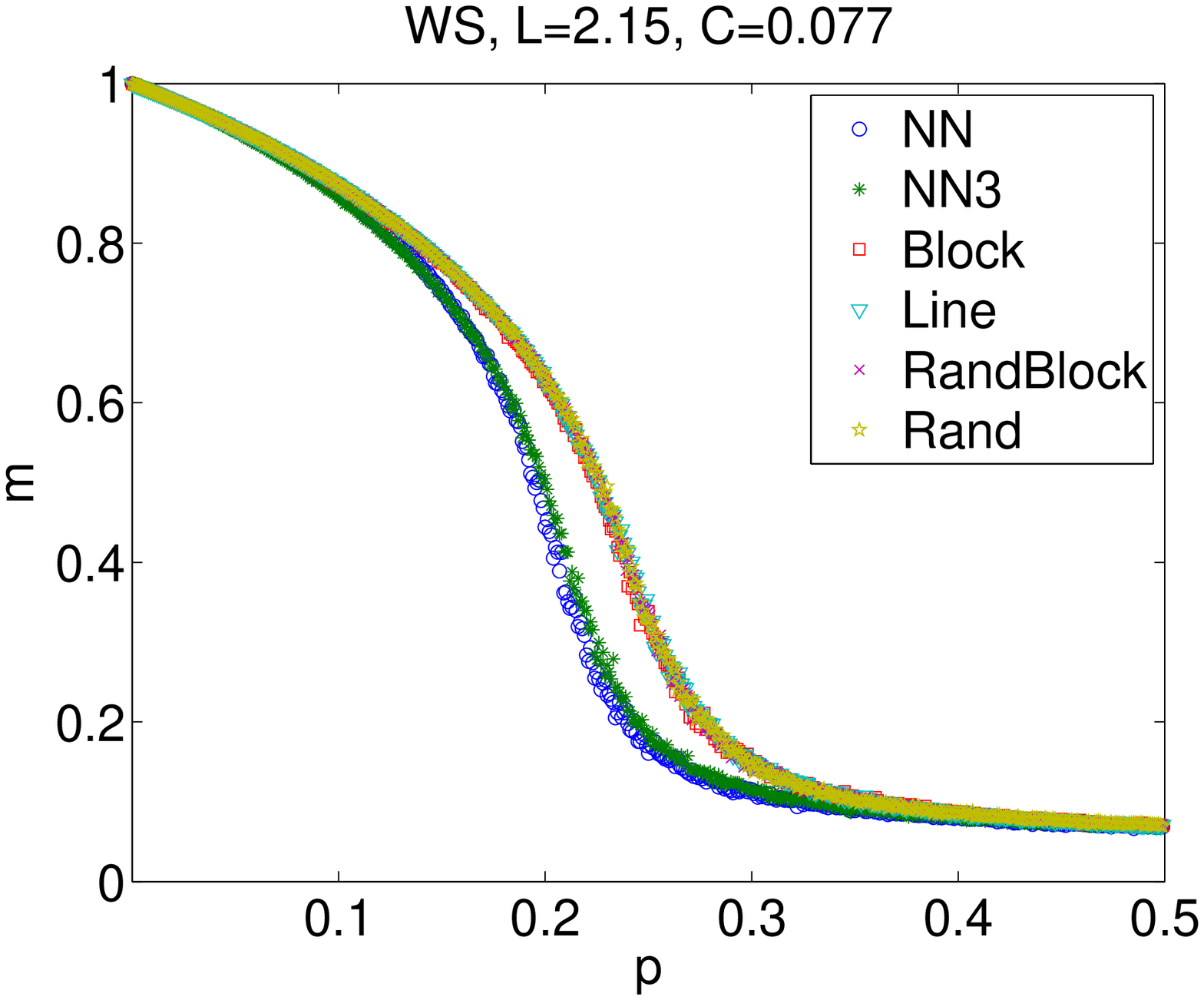} \
\includegraphics[scale=0.22]{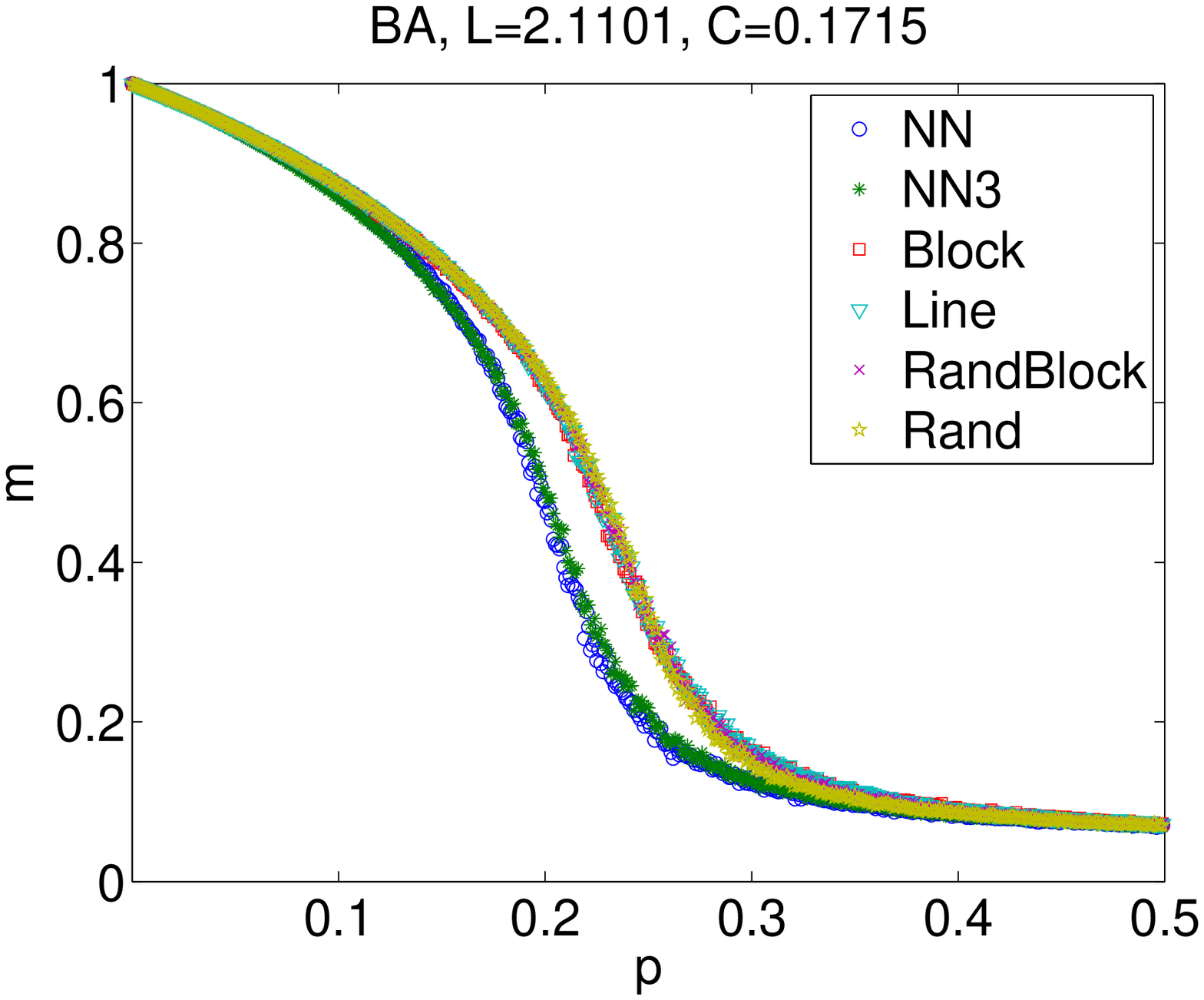} 
\caption{Magnetization $m$ as a function of the independence factor $p$ on six different networks of size $N=233$. In the top row results on three real Twitter ego networks~\cite{McA:Les:2012} are presented whereas in the bottom row results on three artificial networks are shown. The Twitter network IDs in the titles of the plots correspond to file names (`ego' node IDs) in the Twitter dataset taken from \texttt{https://snap.stanford.edu/data/egonets-Twitter.html}~\cite{McA:Les:2012}. In the bottom row results on three model networks are shown: a Watts-Strogatz network with the average degree $K=4$ and the rewiring probability $\beta=0.035$ (bottom left), a WS network with $K=18$ and $\beta=1$ (bottom middle) and a Barab\'asi-Albert one with the number of new edges $M=M_0=10$ (bottom right).\label{fig_real}}
\end{figure}

To check if our predictions about the relative path length are correct, we decided to investigate the problem on several networks of the same size, including real Twitter networks. We took Twitter data from the Stanford Large Network Dataset Collection available at~\url{https://snap.stanford.edu/data/egonets-Twitter.html}~\cite{McA:Les:2012}, because it includes about 1000 different networks with a broad spectrum of diverse characteristics. It was relatively easy to find in the dataset networks of the same size, but with different average path lengths and/or clustering coefficients. Thus the dataset was well suited for testing our hypothesis about the path lengths. Magnetization $m$ as a function of the independence factor $p$ on six different networks of size $N=233$ is presented in Fig.  \ref{fig_real}. In the top row results on three real Twitter networks \cite{McA:Les:2012} are presented. Networks in the left and middle top panels have almost identical path length $l$ but different clustering coefficient $C$. On the other hand, middle and right networks have almost identical clustering coefficient but slightly different path length. It seems that results for all methods overlaps the best on the right network, which has the shortest path length. Simultaneously, it seems that results on left and middle networks are the most similar to each other, i.e. path length $l$  is more significant than clustering coefficient $C$ in determining if all mapping methods will give the same result or not. However, because the differences between properties that we take into account (i.e. $l$ and $C$) do not vary much from network to network, for all three Twitter networks almost all methods collapse into a single curve and only $NN$ and $NN3$ slightly deviate from others. In the bottom row of Fig.~\ref{fig_real} results on three artificial networks are presented. We took two Watts-Strogatz networks (left and middle plots) and a Barab\'asi-Albert one. In case of the WS network with the clustering coefficient similar to Twitter networks and much longer path length $l$, each mapping gives a completely different result (bottom left). On the other hand, if the path lengths are similar to those of real networks, for both WS and BA models we observe already known behavior  - most methods collapse into one curve and only $NN$ and $NN3$ differ slightly from the others. Thus, this feature does not depend on the network topology and the clustering coefficient, which again confirms that the path length is the significant property for the investigated problem.

From obtained results we also conclude that nor the average degree neither the degree distribution are significant for the investigated problem. We realize that there are more properties of the real networks that could be taken into account, but our prediction about the importance of the relative path length seems to be reasonable also from the statistical physics point of view. It should be noted that for all networks considered in this paper two methods collapse into a single curve - RandBlock and Rand and overlap mean-field results presented in \cite{Nyc:Szn:Cis12}. This can be easily understood in the case of Rand method - regardless of the topology, neighbors are randomly chosen. In the case of RandBlock, obtained results are less obvious. However, this method introduces interactions with the infinite range. We know from statistical physics that the mean field approach should give exact results in the case of infinite interactions. Following this reasoning we can also understand why, with decreasing path length, results for all mapping methods approaches the mean field results - the relative interaction range increases. Another phenomena that can be understood on this basis is the fact that $NN$ method differs the most from the $Rand$ and $Line$ is the most similar. $NN$ method has relatively the shortest range of interactions and $Line$ much larger.

\begin{figure}
\centering
\includegraphics[scale=0.4]{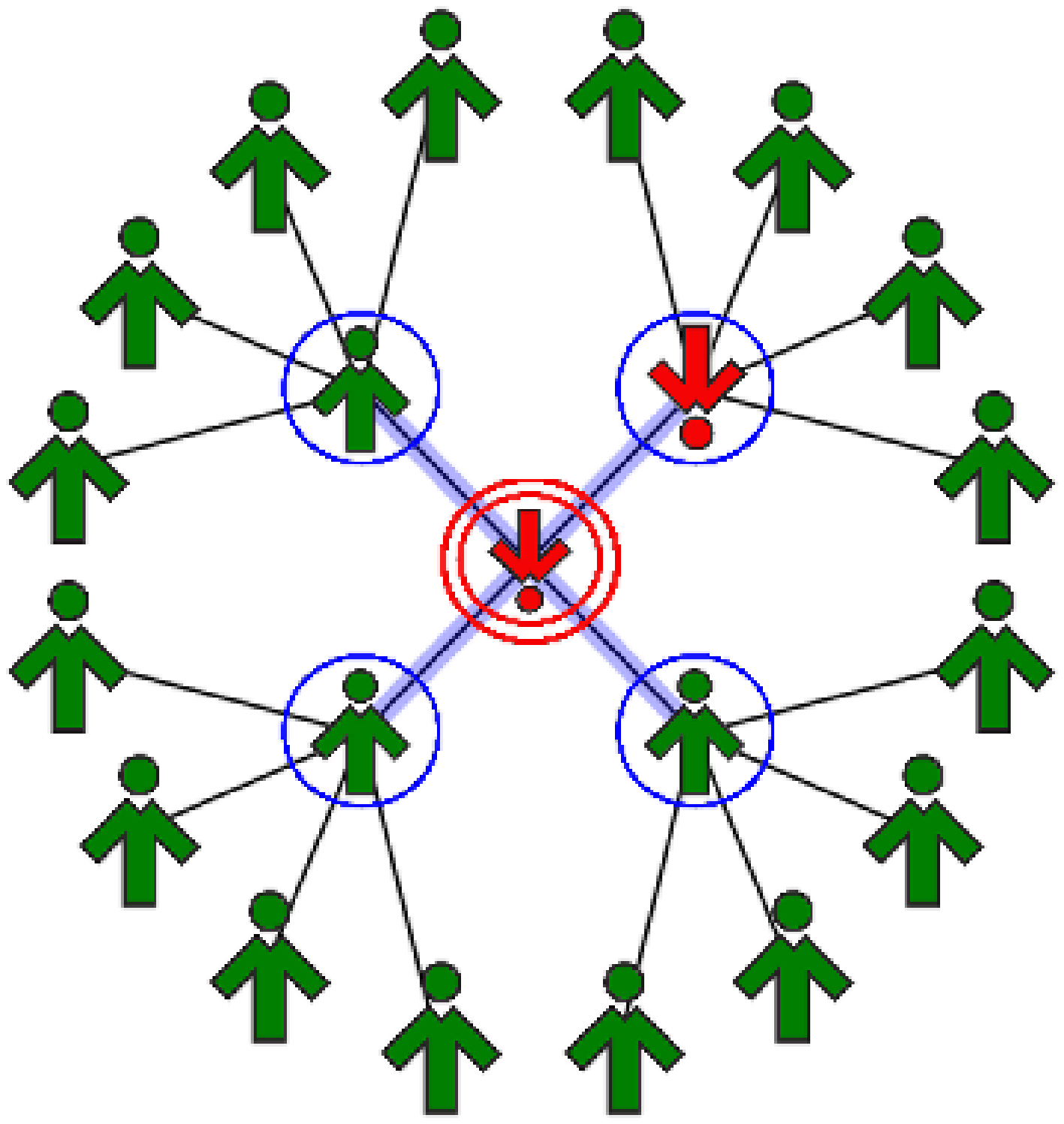} \
\includegraphics[scale=0.4]{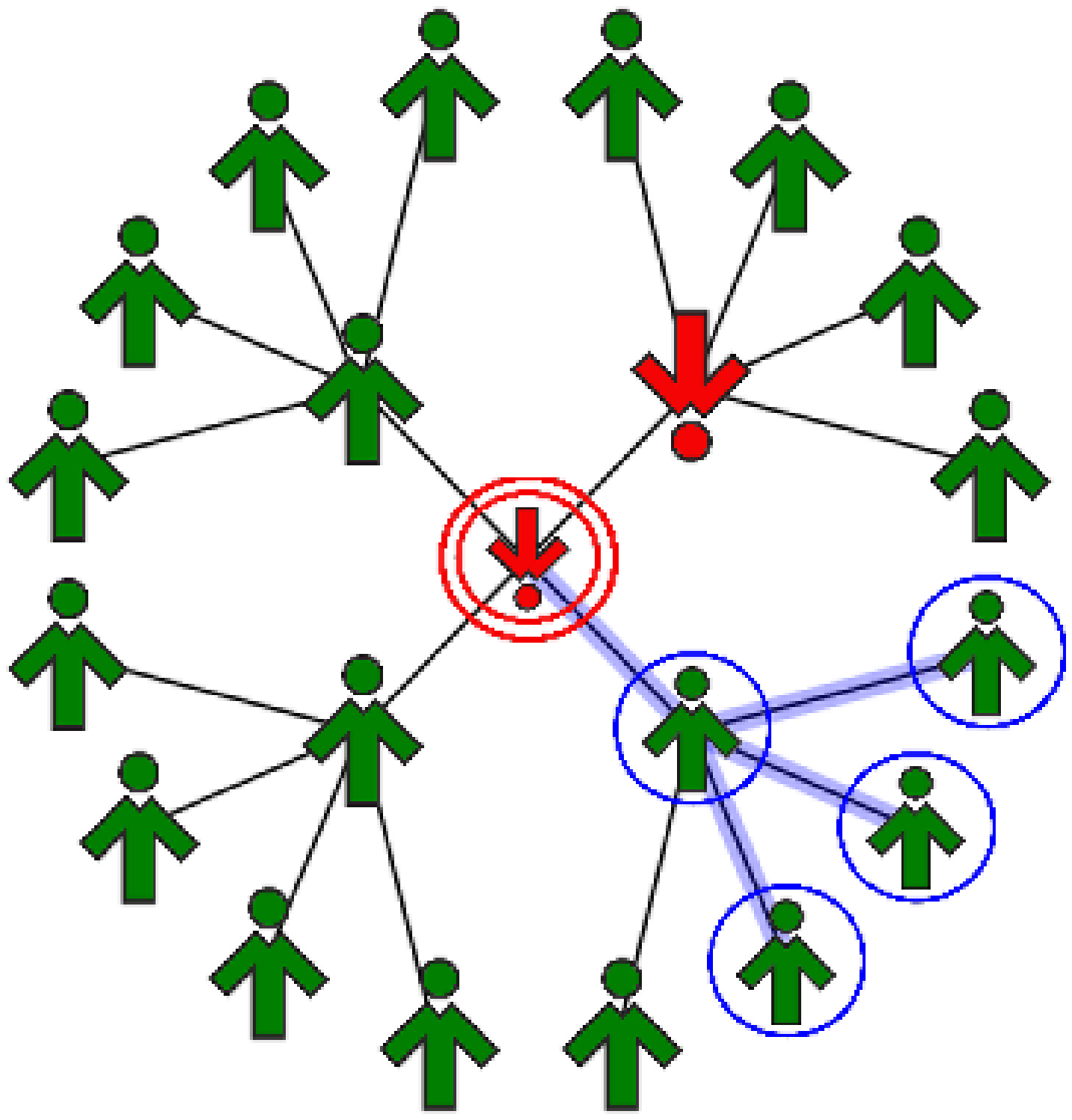} \\
\includegraphics[scale=0.4]{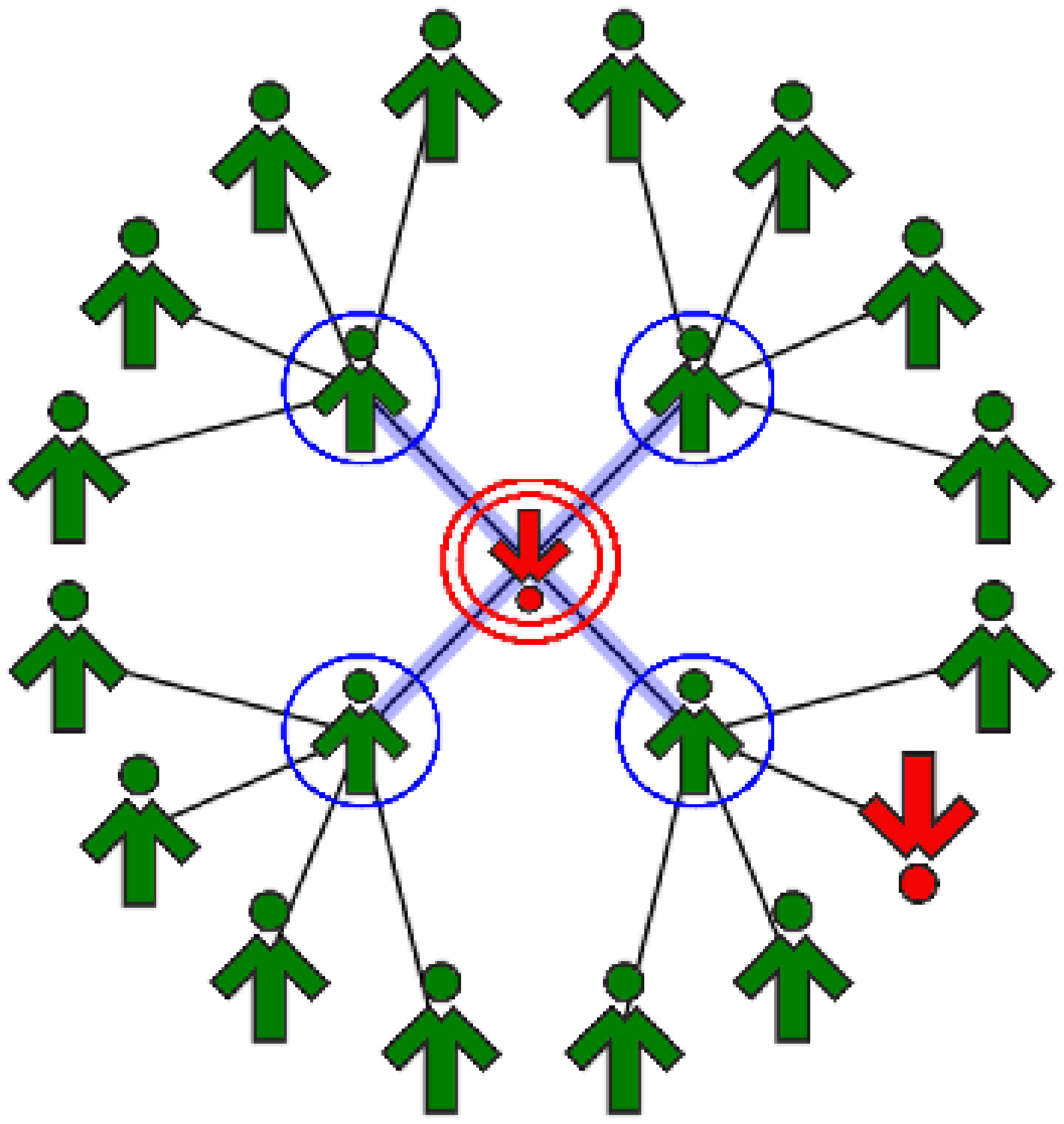} \
\includegraphics[scale=0.4]{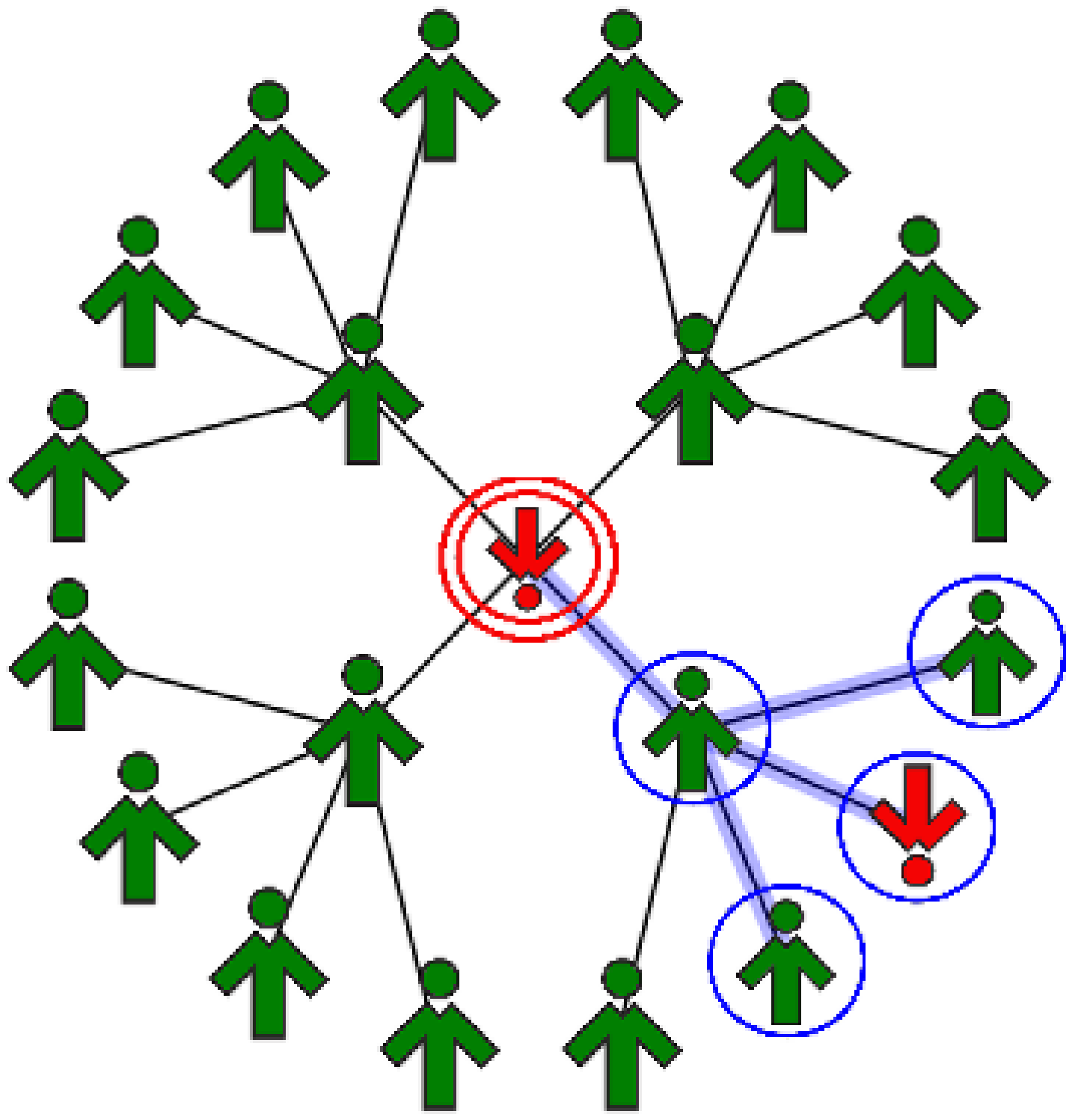} \
\caption{(Color online) A schematic representation of our model at an early stage of a simulation on a Bethe lattice with the coordination number $k =5$ and the root node having $ k-1=4$ neighbors. There are only two not-adopted spinsons (red) in the system and one of them (in double red circle) is exposed to group pressure (group members are marked with blue circles). Top row: if the other red spinson resides in the closest neighborhood of the target one, the $NN$ method (left) gives much higher probability of finding a non-unanimous group of influence than the $Block$ one (right). Bottom: if the not-adopted spinson is at the second level, the $Block$ method yields higher probability than $NN$ (in $NN$ case the probability is zero).  \label{influence_group trees} }
\end{figure}

The differences between methods may be also explained, at least qualitatively, in terms of probabilities of finding non-unanimous influence groups. 
For the sake of simplicity let us assume that our network has the topology of a Bethe lattice~\cite{Osti:2012} with the coordination number $k$. For convenience, we took a modified definition of the Bethe lattice with the central node having only $k-1$ neighbors (Fig.~\ref{influence_group trees}). Thus, the central node has $ k-1$ neighbors in its closest neighborhood, $( k-1)^2$ agents at the second level of its ego graph, and in general  $( k-1)^d$ nodes at distance $d$.
Now, let us consider our model at an early stage of a simulation. Let us assume that there are only two spinsons including the central one in the ``down'' state due to independence and that the central spinson has been chosen again in a basic Monte Carlo event (it will be referred as the  \textbf{target} spinson in the following). However this time it is not independent, i.e. it is exposed to the group pressure. Since most of the spinsons are in the ``up'' state, the system has a natural tendency to reduce disorder due to conformity. Nevertheless, we can ask the question whether there are significant differences between the methods in maintaining disorder in the system. In other words we can check if the methods differ in the probabilities of finding a non-unanimous group of influence in this situation.

To this end, we can consider configurations with the other ``down'' spinson residing at different levels of the target's ego graph. We start with the ``down'' spinson being in the nearest neighborhood of the target node. In this case we expect the $NN$ method to give the highest probability to build a group the ``down'' spinson belongs to. The reason is simple: this is the only method which operates exclusively in the nearest neighborhood of the target spinson. Thus we draw 4 agents out of $ k-1$ to form the group and it is very likely that the ``down'' spinson will belong to the group (see top left plot of Fig.~\ref{influence_group trees} for a schematic representation). The probability of finding a non-unanimous group is slightly smaller in $NN3$ case, because here only 3 drawings out of $k-1$ from the closest neighborhood are allowed. The $Line$ and $Block$ methods require only one drawing from the first level (top right plot of Fig.~\ref{influence_group trees}). Hence it is less likely to hit the ``down'' spinson. Finally, $RandBlock$ and $Rand$ algorithms yield the smallest probabilities, because they operate on the whole network rather than in the close neighborhood of the target node.

Since the problem at hand is nothing but a variation of an urn problem~\cite{Joh:Kot:1977}, we can actually calculate for each method the probability of finding a non-unanimous group of influence. To focus our attention we set  $ k  =9$ and the system size $N=4680$ meaning that the ego graph of the target spinson consists of 4 levels. In the case of the $NN$ method the number of all possible $q$-panels is just the number of 4-combinations selected from the nearest neighborhood of the target, 
\begin{equation}
|\Omega | = { k-1 \choose 4} = {8 \choose 4} = \frac{8!}{4!4!}=70.
\end{equation} 
The not adopted agent at the first level has to belong to each non-unanimous group. The other three members are selected from seven ``up'' spinsons residing at that level. Thus, the number of non-unanimous groups in the $NN$ method is equal to
\begin{equation}
|NU | = {1 \choose 1}{7 \choose 3} = \frac{7!}{3!4!}=35.
\end{equation} 
This yields the following probability of finding a non-unanimous group:
\begin{equation}
P^{(1)}_{NN} = \frac{|NU|}{|\Omega|} = 0.5.
\end{equation} 
The superscript $(1)$ in the last expression indicates the level the other ``down'' spinson belongs to. 
In the $NN3$ method we select three agents from the first level, then pick one out of them and draw one spinson from its neighbors at the next level. This gives
\begin{equation}
|\Omega | = {8 \choose 3}{3 \choose 1}{8 \choose 1} = 1344
\end{equation} 
possible influence groups. Once the ``down'' spinson is chosen, we select two others from the first level, then pick one of the members of the group and add one of its neighbors to the group. The number of all  possibilities is in this case given by
\begin{equation}
|NU | = {1 \choose 1}{7 \choose 2}{3 \choose 1}{8 \choose 1} = 504.
\end{equation} 
Hence,
\begin{equation}
P^{(1)}_{NN3} = \frac{|NU|}{|\Omega|} = 0.375.
\end{equation} 
Similar analysis leads to the following results for the other methods:
\begin{eqnarray}
& & P^{(1)}_{Block} = 0.125,~~P^{(1)}_{Line} = 0.125,\nonumber\\
& & P^{(1)}_{Rand} = 0.85*10^{-3},~~P^{(1)}_{RandBlock} = 0.21*10^{-3}.
\end{eqnarray} 
We see that indeed the $NN$ method gives the highest probability of leaving the target spinson untouched in this case.

If the other ``down'' spinson resides at the second level of target's ego graph, the $Block$ method should give the highest probability of maintaining disorder in the system, because it consists of drawings mostly from that level (bottom row of Fig.~\ref{influence_group trees}). Again, we can calculate the corresponding probabilities to get:
\begin{eqnarray}
& & P^{(2)}_{Block} = 0.046,~~P^{(2)}_{Line} = 0.015,~~P^{(2)}_{NN3} = 0.015,\nonumber\\
& & P^{(2)}_{Rand} = 0.85*10^{-3},~~P^{(2)}_{RandBlock} = 0.29*10^{-3},~~P^{(2)}_{NN} = 0.
\end{eqnarray} 
The $Block$ method gives indeed the highest probability, followed by $Line$ and $NN3$. The probability for $NN$ is zero, because the method operates only at the first level. It is interesting to note that the probabilities for the second level are much lower than those for the first one. With similar reasoning we can show that in general the farther the distance between the two ``down'' spinsons, the smaller the chance to maintain disorder, i.e. to let the target spinson unchanged.  As a consequence, only the state of the two closest levels is actually significant for the evolution of target's opinion. For this reason the methods $NN$ and $NN3$, followed by $Block$ and $Line$ destroy the order in the systems the fastest, i.e. for relative small values of the independence parameter $p$. 

Note that the above conclusion is in accordance with our simulation results shown in Figs.~\ref{fig_ba},~\ref{fig_ws} and~\ref{fig_real}. Thus, although a Bethe lattice resembles ego graphs of model and real networks on average only, the reasoning remains the same - if the ``down'' agents are sparse, the $NN$ and $NN3$ methods yield the highest probability to maintain disorder in the system.  

From our simulations it follows that in networks with  short relative average path lengths the differences between all but $NN$ and $NN3$ are negligible. A short path length means usually that the ego networks of all agents are rather ``flat'', i.e. most agents reside at very few levels. In this case already the second and next levels of an ego graph are highly populated, leading to negligible probabilities of finding non-unanimous groups at the beginning  of a simulation for methods operating mainly beyond the first level. Hence, on such networks the results for four methods ($Block$, $Line$, $RandBlock$ and $Rand$) are essentially indistinguishable. The $NN$ and $NN3$ methods deviate slightly from others giving a bit smaller critical values of the independence $p$, i.e. destroying the order a bit faster (see Fig.~\ref{fig_ba}, bottom row of Fig.~\ref{fig_ws} and most of the plots in Fig.~\ref{fig_real} for further reference).

\section{Conclusions}
From physical point of view it is always an interesting question how details at the microscopic scale manifest at the macroscopic level. In the field of opinion dynamics such 
a macroscopic quantity is the opinion, defined in a case of binary models as the magnetization $m$. In this paper we examine six models that differ only in the way of selecting a group of influence but the size of this group remains fixed. Therefore there are no differences between models on the complete graph. For other topologies methods for which range of interaction is shorter tend to show the lower critical value of the independence factor $p$ below which $m>0$ and above $m=0$. Only two methods, $RandBlock$ and $Rand$, give exactly the same results on all topologies and overlap MFA result found in \cite{Nyc:Szn:Cis12}. Remaining four methods give different results and differences between methods increase with the relative path length, i.e. are the most visible on the regular lattices. With decreasing relative path length the differences between methods vanish. One should notice that the average path length itself does not determine the differences between models, because results are virtually independent on the system size. What determines network properties is the relative path length defined as the ratio between the average path length of considered network and the average path length of a random graph of the same size $L$ and average degree $\langle k \rangle$.
It should be noted that most of the real-world networks are characterized by relatively short paths \cite{alb02} and therefore differences between models should be negligible. 

We believe that our results contribute also to the discussion on differences between inflow and ouflow dynamics. As noted by Dietrich Stauffer: \emph{The crucial difference of the Sznajd model compared with voter or Ising models is that information flows outward: A site does not follow what the neighbors tell the site, but instead the site tries to convince the neighbors} \cite{Stau:02}. Of late, a debate on whether inflow dynamics is different from outflow dynamics has emerged \cite{Szn:Kru:06,Cas:Pas:11,Roy:Bis:Sen:14}. Our findings indicate that not the direction of the information flow itself but the range of interactions is important, what coincides with results obtained by Castellano and Pastor-Satorras \cite{Cas:Pas:11}.  It is worth to notice that some of rules, investigated here, may be viewed as inflow and other as outflow dynamics. In particular, the $NN$ method corresponds to the inflow dynamics. On the other hand, the $Block$ method was inspired by the two dimensional  and the $Line$ model by the one dimensional outflow dynamics. Therefore both can be viewed as outflow dynamics. Both outflow rules ($Block$ and $Line$) give the same results on scale-free and real Twitter networks, whereas the inflow rule $NN$ gives lower value of the critical value of $p$. However, it seems that the critical value of $p$ increases with the relative range of interactions and therefore it is understandable that $NN$ (inflow) rule gives the lowest value of $p$. So perhaps one should not think about the direction (in or out) of the information flow itself but the range of interactions, what would coincide with results obtained by Castellano and Pastor-Satorras \cite{Cas:Pas:11}. Summarizing, indeed inflow and outflow dynamics give different results but the reason is simply the difference in the range of interactions.

As already mentioned in the introduction, the main motivation for this paper was the remark by Macy and Willer that \emph{there was a little effort to provide analysis of how results differ depending on the model designs} \cite{Mac:Wil:02}. In the context of the problem posed here, it would seem that the structure of the group of influence may be important from the social point of view. However, as we have shown in the case of many complex networks, including BA and real networks, the importance of the group structure is often negligible.

In this paper we considered only static networks (not changing in time), which is a common approach while studying dynamical processes like opinion spreading or diffusion of innovation. However, the characteristics of many real networks evolve in time and there is more and more data available on temporal networks~\cite{Eag:Pen:2006}. If the changes take place at time scales comparable to those of  studied processes, the temporal heterogeneities in such networks may lead to big differences in the dynamics  of the processes, even if the networks appear similar from the static perspective~\cite{Pan:Sar:2011}. Thus, it could be worth to check what is the impact of the group structure in models put on top of real temporal networks. This issue will be addressed in one of the forthcoming papers.  

\section*{Acknowledgments} This work was supported by funds from the National Science Centre (NCN) through grant no. 2013/11/B/HS4/01061.

\end{document}